\documentclass{article}
\usepackage{natbib}
\usepackage{graphicx}
\usepackage{booktabs}
\usepackage{multirow}
\usepackage{soul}
\usepackage{lipsum}
\usepackage{amsmath}
\usepackage{textcomp}
\usepackage{xcolor}
\usepackage{geometry}
\usepackage{hyperref}
\usepackage{listings}
\usepackage{arydshln}
\usepackage{geometry}
\setcitestyle{authoryear,open={(},close={)}} 

\newgeometry{vmargin={15mm}, hmargin={12mm,12mm}}   %

\begin{document}
\newcommand{\Equref}[1]{Equation~\ref{#1}}
\newcommand{\equref}[1]{Equation~\ref{#1}}
\newcommand{\Secref}[1]{Section~\ref{#1}}
\newcommand{\secref}[1]{Section~\ref{#1}}
\newcommand{\dashsecref}[2]{Sections~\ref{#1}--\ref{#2}}
\newcommand{\Defref}[1]{Definition~\ref{#1}}
\newcommand{\defref}[1]{Definition~\ref{#1}}
\newcommand{\Figref}[1]{Figure~\ref{#1}}
\newcommand{\figref}[1]{Figure~\ref{#1}}
\newcommand{\dashfigref}[2]{Figures~\ref{#1}--\ref{#2}}
\newcommand{\Tabref}[1]{Table~\ref{#1}}
\newcommand{\tabref}[1]{Table~\ref{#1}}

\title{DebiasedDTA: A Framework for Improving the Generalizability of Drug-Target Affinity Prediction Models}

\author{R{\i}za \"{O}z\c{c}elik\,$^1$,
Alperen Ba\u{g}\,$^{1,+}$,
Berk At{\i}l\,$^{1,+}$,
Melih Barsbey\,$^{1,+}$,
Arzucan \"{O}zg\"{u}r\,$^{1,*}$, 
Elif Ozkirimli\,$^{2,*}$}

\date{}

\maketitle
\noindent$^1$Department of Computer Engineering, Bo\u{g}azi\c{c}i University, \.{I}stanbul, Turkey and \\
$^2$Data and Analytics Chapter, Pharma International Informatics, F. Hoffmann-La Roche AG, Switzerland \\
$^+$These authors contributed equally to the work.\\
$^\ast$To whom correspondence should be addressed.

\begin{abstract}
Computational models that accurately predict the binding affinity of an input protein-chemical pair can accelerate drug discovery studies. These models are trained on available protein-chemical interaction datasets, which may contain dataset biases that may lead the model to learn dataset-specific patterns, instead of generalizable relationships. As a result, the prediction performance of models drops for previously unseen biomolecules, \textit{i.e.} the prediction models cannot generalize to biomolecules outside of the dataset. The latest approaches that aim to improve model generalizability either have limited applicability or introduce the risk of degrading prediction performance. Here, we present DebiasedDTA, a novel drug-target affinity (DTA) prediction model training framework that addresses dataset biases to improve the generalizability of affinity prediction models. DebiasedDTA reweights the training samples to mitigate the effect of dataset biases and is applicable to most DTA prediction models. The results suggest that models trained in the DebiasedDTA framework can achieve improved generalizability in predicting the interactions of the previously unseen biomolecules, as well as performance improvements on those previously seen. Extensive experiments with different biomolecule representations, model architectures, and datasets demonstrate that DebiasedDTA can upgrade DTA prediction models irrespective of the biomolecule representation, model architecture, and training dataset. Last but not least, we release DebiasedDTA as an open-source python library to enable other researchers to debias their own predictors and/or develop their own debiasing methods. We believe that this python library will corroborate and foster research to develop more generalizable DTA prediction models.
\end{abstract}

\section{Introduction}

Proteins and chemicals interact with each other by following fundamental physicochemical principles, and a complete understanding of these principles would allow the accurate prediction of protein-chemical interactions.
Without such an understanding, machine learning approaches that rely on the available knowledge space of large interaction datasets can help to rapidly identify high-affinity protein-chemical pairs in the vast combination space by learning affinity patterns from measurements for millions of protein-chemical pairs. However, a representative sampling of the entire combination space is impossible or infeasible, and thus the available datasets explore only portions of this space \citep{mestres2008data}. Therefore, machine learning algorithms that are trained on these available datasets are prone to learn some spurious relationships that are specific to those datasets and that are not generalizable to 
whole combination space (e.g., the presence of sulfur may separate most actives and inactives of a target in a dataset, even though it is not a major determinant of binding affinity in general) \citep{bietz2015discriminative,chaput2016benchmark,wallach2018most,sieg2019need}. Previous work suggests that these potentially misleading spurious patterns may be relatively easier for prediction models to pick up and can reduce their generalizability to novel proteins and chemicals, for which the learned patterns are non-applicable \citep{chen2019hidden,tran2020lit,yang2020predicting,boyles2020learning,ozccelik2020chemboost}. The problem of overcoming dataset biases in generalization performance has been studied under various concepts in machine learning literature, including but not limited to ``learning under distribution shifts'', ``domain generalization'', and ``out-of-distribution generalization'' \citep{kouwReviewDomain2021}. The fact that models can pick up on spurious, simple relationships that would lead their predictions to suffer in terms of generalizability has been observed in the wider machine learning field as well \citep{xiao2020noise,shah2020pitfalls,geirhos2020shortcut}.

Recent studies that address the problem of training generalizable prediction models for protein-chemical interactions have designed train-test splits with dissimilar biomolecules so that the dataset biases are less rewarding on validation and test sets  \citep{pahikkala2014toward,guney2017reproducible,wallach2018most,tran2020lit}. However, these approaches fail to improve predictive performance for novel biomolecules and produce inaccurate estimates of the performance on dissimilar test sets \citep{sundar2019effect}. Other studies adopted data augmentation to address the problem, but either their applicability was limited by the number of available 3D structures \citep{scantlebury2020dataset} or they were non-scalable to datasets with a large number of proteins \citep{sundar2020using}. Furthermore, all of these studies approached the problem for the ligand-based virtual screening setting and treated the problem as a binary classification problem (active/inactive). To the best of our knowledge, there is no study that proposes a framework for enhancing drug-target affinity (DTA) prediction on novel biomolecules in a regression setting.  Therefore, drug-target affinity prediction on unseen biomolecules remains a significant challenge in the field, especially in drug discovery against drug targets for rare diseases or in identifying novel chemical moieties. 

In the machine learning literature at large, various approaches have been developed to address the problem of generalization under distribution shift \citep{kouwReviewDomain2021}. Example approaches include modifying the training dataset, feature representations, and/or the inference procedure to match the characteristics of the test distribution \citep{rahmanCorrelationawareAdversarial2019},
changing the optimization procedure to improve worst-case performance in the space around the training distribution \citep{sagawaDistributionallyRobust2020}, and learning causal/invariant relationships within the data under the assumption that such relationships are robust across different distributions \citep{arjovskyInvariantRisk2020,shenOutOfDistributionGeneralization2021}. Here, we follow the latter path since protein-chemical interactions are governed by physicochemical laws that can be expected to be robust across datasets.

In this paper, we propose DebiasedDTA,  a novel model training framework to address dataset biases and improve the performance of DTA prediction models for novel biomolecules. The method utilizes the causal nature of the protein-chemical interaction mechanism to obtain a predictor attuned to this invariant relationship.
DebiasedDTA is model- and representation-agnostic; in that, it can be combined with any DTA prediction model that allows reweighting of the instances and thus can find a wider application range than approaches proposed for ligand-based virtual screening.

The DebiasedDTA training framework comprises two stages, which we call ``guide" and ``predictor" models. The guide learns a reweighting of the dataset such that a model trained thereupon can learn a robust relationship between biomolecules and binding affinity instead of spurious associations. The predictor then uses the weights produced by the guide and progressively weights the training data during its training to generalize well to unseen biomolecules.

We experiment with two guide and five predictor variants to evaluate the proposed DebiasedDTA training framework. The results show that the proposed model training framework can increase robustness to dataset biases and enhance the out-of-distribution  prediction performance of DTA models with various drug-target representations. Furthermore, the improvement is  observed  not only for  novel biomolecules but also for the previously encountered ones.

The problem of overcoming dataset biases to achieve out-of-distribution generalization is a notoriously difficult one, and previously proposed methods still fail to consistently outperform basic empirical risk minimization in generalization performance within computational drug discovery \citep{jiDrugOODOutofDistribution2022} and in machine learning at large \citep{gulrajaniSearchLost2020,kohWILDSBenchmark2021-arxiv}. Therefore our work is a step towards achieving successful DTA prediction models for novel, as well as previously seen biomolecules. To foster future research in this quest, we release pydebiaseddta: an easy-to-use python library where researchers can test their own guide and/or predictors, develop new debiasing techniques, and benchmark the generalizability of their models across datasets. The installation and usage instructions are available on GitHub: \url{https://github.com/rizaozcelik/pydebiaseddta}

\begin{figure*}[t]
    \centering
    \includegraphics[width=\textwidth]{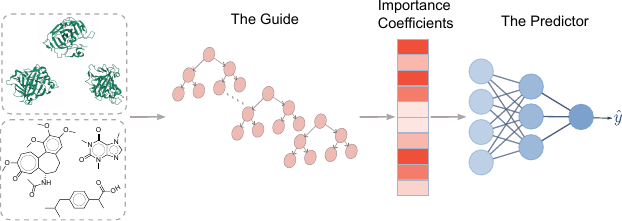}
    \caption{DebiasedDTA. DebiasedDTA training framework improves the generalizability of DTA prediction models on novel
biomolecules by targeting dataset biases. In DebiasedDTA, the “guide” models learn a weighting of the training set, called
importance coefficients, such that the protein-chemical affinities that cannot be predicted merely through dataset biases have
higher weights. The weights are determined as the median squared error of each sample resulting from training the guides using
10-fold cross-validation on the training set. Here, we experiment with ID-DTA and BoW-DTA to target biomolecule word
and identity-driven biases with the guides. The “predictors”, use the importance coefficients to prioritize training samples
and attribute higher importance to the instances that challenge the guides. The biomolecule representation of the predictors
can take any form and we experiment with five models (DeepDTA, BPE-DTA, LM-DTA, GraphDTA, and mGraphDTA) to
evaluate DebiasedDTA with sequence, graph, and pre-trained representations. The experiments show that the DebiasedDTA
training framework can enhance the performance of all models on unseen biomolecules, as well as the known ones.}
    \label{fig:debiaseddta}
\end{figure*}

\section{Materials and Methods}
\subsection{DebiasedDTA}
DebiasedDTA consists of two components, which we refer to as ``guide'' and ``predictor''. The guide aims to weight the training set instances in a way that facilitates learning the invariant relationships between biomolecules and their affinity, instead of dataset-specific spurious relationships. The predictor, on the other hand, can be any DTA prediction model that can use instance weighting in its loss function. These DTA prediction models include, but are not limited to, deep learning models, kernel machines, and tree-based ensemble models.

Given the high-dimensional nature of biomolecule inputs under commonly used representations, and the difficulty of collecting experimental data on affinity for all possible protein-chemical pairs, training datasets contain narrow regions of the potential interaction space and they are highly likely to contain spurious associations that can be picked up by supervised models. Consequently, these models would fail to generalize to unseen data when the test distribution does not contain these spurious associations. The current method aims to address this problem by increasing the weights of observations that are likely to contain unique information that is not captured by such spurious associations and training the DTA prediction model with these new weights.

Let $x_p, x_l$ denote a protein-ligand pair (with an arbitrary representation) and let $y$ denote their affinity. The aim of any supervised DTA prediction task is to estimate $p(y|x_p, x_l)$. Importantly for our study, as described in the previous section, the conditional distribution $ p(y|x_p, x_l)$ can be assumed to be \textit{invariant} between any training or test distributions since this corresponds to a mechanical/causal relationship. This corresponds to the \textit{independent causal mechanisms} assumption from the causal inference literature \citep{petersElements2017}. %

Let $x^h_p, x^h_l$ be some statistics or features of the pair $x_p, x_l$ (such as the non/existence of patterns of chemical moieties or amino acids) that are sufficient to predict DTA, such that $p(y|x_p, x_l) = p(y|x^h_p, x^h_l)$, latter also being invariant between datasets. Also, let $x^s_p, x^s_l$ be other, lower-dimensional statistics of $x_p, x_l$, such as the non/existence of a single atom or amino acid in the biomolecules. As opposed to the generative distribution of drug-target affinities $p$, let $p_S$ and $p_T$ denote the probability distributions for a hypothetical training and test set pair. %
If the training data distribution $p_S$ conforms to the graphical model
$$y \leftarrow x^h_p, x^h_l \rightarrow x^s_p, x^s_l,$$ this would induce a spurious relationship between $y$ and $x^s_p, x^s_l$. Then, the learned posterior distribution $\hat{p}_S$ would end up satisfying $\hat{p}_S(y|x_p, x_l) = \hat{p}_S(y|x^s_p, x^s_l).$ This estimated relationship is not \textit{robust} to distribution shift, in the sense that it would not lead to correct decisions in a test distribution $p_T$ where this spurious relationship is either nonexistent%
, i.e. $y \leftarrow x^h_p, x^h_l \not\rightarrow x^s_p, x^s_l,$
or changed in any other way.

To combat this, in addition to the causal invariance, the present method assumes that (i) the spurious $y\leftrightarrow x^s_p, x^s_l$ relationship can be learned with a low-complexity model, and (ii) for a given non-generalizable spurious relationship, there are rare observations in the data set that do not conform to this relationship. Therefore, after training with a simple learner with a subset of the training data, if the likelihood of the prediction for a held-out training instance is low, i.e. its error is high,  the weight of this instance is increased (by the \textit{guide}, see below for details). This effectively modifies the training distribution by increasing the presence of samples that do not conform to these spurious patterns, thereby helping the \textit{predictor} to estimate $\hat{p}_S$ such that, ideally  $$\hat{p}_S(y|x_p, x_l) = \hat{p}_S(y|x^h_p, x^h_l) \neq \hat{p}_S(y|x^s_p, x^s_l).$$ This implies better generalizability, since $p_S(y|x^h_p, x^h_l) = p(y|x^h_p, x^h_l)$ as argued above. Finally, note that the estimation of $p(y|x^h_p, x^h_l)$ enables the use of the predictor on novel biomolecules since a novel molecule can still contain previously encountered statistics.

\subsubsection{The Guide}
The guides in DebiasedDTA target the low-complexity spurious relationships in the data and, therefore should have a limited learning capacity. So, we design two weak learners with simple biomolecule representations: an identifier-based model, ID-DTA, and a biomolecule word-based model, BoW-DTA. ID-DTA is motivated by the fact that the mere use of random biomolecule identifiers can produce high-achieving models for similar test sets \citep{ozccelik2020chemboost}, and thus, can capture the said dataset biases. ID-DTA vectorizes the chemicals and proteins in the training set with one-hot-encoded vectors and reserves two dimensions to represent unseen chemicals and proteins during prediction.

BoW-DTA is inspired by natural language inference studies in which the use of certain words is exploited by the prediction models to predict semantic labels \citep{gururangan2018annotation,poliak-etal-2018-hypothesis}.  We create BoW-DTA to investigate a similar relationship between biomolecule sequences and affinity scores. BoW-DTA segments biomolecule sequences into biomolecule words with the Byte Pair Encoding \citep{sennrich2015neural} algorithm and adopts the bag-of-words approach to vectorize the biomolecules. Both ID-DTA and BoW-DTA concatenate biomolecule vectors to represent biomolecule pairs and use decision tree regression for prediction, as decision trees have limited learning capacity and yet can learn non-linear relationships. The python framework we publish as an open-source software allows other such guide mechanisms to be proposed and benchmarked by future research.

We leverage the guides to identify protein-chemical pairs that bear more information about the mechanisms of protein-chemical binding. We hypothesize that if the guides, models designed to learn misleading spurious patterns, perform poorly on a protein-chemical pair, then the pair is more likely to bear generalizable information on binding and deserves higher attention by the DTA predictors. We adopt 5-fold cross-validation to measure the performance of a guide on the training interactions. First, we randomly divide the training set into five folds and construct five different mini-training and mini-validation sets. We train the guide on each mini-training set and compute the squared errors on the corresponding mini-validation set. One run of cross-validation yields one squared-error measurement per protein-chemical pair as each pair is placed in the mini-validation set exactly once. In order to better estimate the performance on each sample, we run the 5-fold cross-validation 10 times and obtain 10 error measurements per sample. We compute the median of the 10 squared errors and call it the ``importance coefficient" of a protein-chemical pair, after normalization by their sum. The importance coefficients guide the training of the predictor after being converted into training weights as described in the next section.

\subsubsection{The Predictor} \label{sec:predictor}
In the DebiasedDTA training framework, the predictor is the model that will be trained with the training samples weighted by the guide to ultimately predict target protein-chemical affinities. The predictor can adopt any biomolecule representation but has to be able to weight the training samples during training to comply with the weight adaptation strategy proposed in DebiasedDTA.

The proposed strategy initializes the training sample weights to 1 and updates them at each epoch such that the weight of each training sample converges to its importance coefficient at the last epoch. When trained with this strategy, the predictor attributes higher importance to samples with more information on binding rules (\textit{i.e.} samples with higher importance coefficient) as the learning continues. Our weight adaptation strategy is formulated as 

\begin{equation}
    \label{eq:bias_decay}
    \vec{w}_e = (1 - \frac{e}{E}) + \vec{i} \times \frac{e}{E},   
\end{equation}

\noindent where $w_e$ is the vector of training sample weights at epoch $e$, $E$ is the number of training epochs, and $\vec{i}$ is the importance coefficients vector. Here, $e/E$ increases as the training continues, and so does the impact of $\vec{i}$, importance coefficients, on the sample weights.

We use five DTA affinity prediction models to evaluate DebiasedDTA on different biomolecule representations: DeepDTA \citep{ozturk2018deepdta}, BPE-DTA, LM-DTA, GraphDTA \citep{nguyen2021graphdta}, and MGraphDTA \citep{yang2022mgraphdta}. DeepDTA and BPE-DTA represent chemicals and proteins with SMILES and amino-acid sequences, respectively,  and learn continuous biomolecule representations with stacked convolutions. These models use the same prediction module (a three-layered fully connected neural network) and differ only in the sequence segmentation strategy -- DeepDTA segments sequences into characters and BPE-DTA relies on biomolecule words. LM-DTA, on the other hand, represents chemicals and proteins with the fixed vectors output by pre-trained language models, ChemBERTa \citep{chithrananda2020chemberta} and ProtBERT \citep{elnaggar2020prottrans}, respectively, and adopts a two-layered fully-connected network for prediction. Last, GraphDTA and MGraphDTA use message passing on molecular graphs to represent chemicals and convolutions on amino-acid sequences to represent proteins. However, they adopt different message-passing networks and convolutions of different depths and granularity, resulting in different biomolecule representations. The aforementioned open-source software we publish allows other predictor models to be included and benchmarked within DebiasedDTA.

\subsection{Experimental Setup}
\subsubsection{Datasets}
We test DebiasedDTA on BDB \citep{ozccelik2020chemboost} and KIBA \citep{tang2014making} datasets. KIBA contains 118K affinity measurements of 229 kinase family proteins and 2111 chemicals, and the affinities are reported in terms of KIBA score. KIBA score combines different measurement sources such as $K_d$, $IC_{50}$, and $K_{i}$, and ranges from 1.3 to 17.2 in the dataset, with increasing KIBA scores denoting higher binding affinity.

BDB is a dataset filtered from BindingDB database \citep{liu2007bindingdb} and comprises 31K binding affinity measurements of 490 proteins and 924 chemicals. The binding affinities are recorded in terms of $pK_d$ (see Equation \ref{eq:pkd}) \citep{ozccelik2020chemboost}, which correlates positively with the binding strength and changes between 1.6 and 13.3 in the dataset. Protein diversity is higher in BDB than KIBA as it contains fewer interactions, but more proteins from different families.

\begin{equation}
    \label{eq:pkd}
    pK_d= -\log_{10} (\frac{K_d}{1e9})
\end{equation}

\subsubsection{Experimental Settings}

We evaluate the models with five distinct setups per dataset. Each setup consists of one training, one validation, and four test sets, which are called warm, cold chemical, cold protein, and cold both. The warm test set contains interactions between previously encountered biomolecules, whereas cold chemical test set consists of the interactions between unseen chemicals and known proteins. This test set estimates model performance when new drugs are searched to target existing proteins. The cold protein test set is created similarly and used to evaluate models in scenarios where existing drugs are searched to target a novel protein. Last, the cold both test set is the set of interactions between novel proteins and chemicals and forms the most challenging test set of every setup, as both the proteins and the chemicals are unavailable in the training set. The dissimilarity of the biomolecules in different splits is enforced by a clustering algorithm, which is explained in the appendix.

To tune the hyper-parameters, we train models on the training set of each setup and measure the performance on the corresponding validation set, which contains interactions only between known biomolecules. We pick the hyper-parameter combination that scores the lowest validation average mean squared error to predict the test set interactions.

\subsubsection{Evaluation Metrics}
We evaluate the models with two metrics, concordance index (CI) \citep{gonen2005concordance} and R$^2$. We use CI to evaluate the similarity of the predicted binding affinity ranking of protein-chemical pairs with the expected one. CI is independent of the output range and allows comparisons across datasets. CI is expected to be around 0.5 for random predictions and reaches 1 when two sets of rankings match exactly. 
R$^2$, on the other hand, is a regression metric that measures how much of the variance in the expected labels is explained by the predictions. R$^2$ equals 1 when labels and predictions are the same, is 0 when all predictions are equal to the mean of test labels, and can drop below 0 when the model underperforms a hypothetical test-mean predictor. 

\subsubsection{Comparing DebiasedDTA with A State-of-the-art Approach}
To the best of our knowledge, DebiasedDTA is the first study that tackles the generalizability problem in drug-target interactions in a regression setup, and therefore, no direct benchmark exists. Closely related methods are proposed under a ligand-based virtual screening framework, and  asymmetric validation embedding (AVE) \citep{wallach2018most}, stands as the state-of-the-art debiasing approach in that context. 

AVE is a measure of chemical bias and minimizing AVE over the training set should also minimize the training bias \citep{wallach2018most}. Being developed for a classification task, AVE requires a list of active and inactive chemicals to minimize training set bias. So, affinity scores in BDB and KIBA are converted to binary labels using the thresholds of 7 and 12.1 \citep{ozccelik2020chemboost}, respectively, and AVE is run to debias training sets of BDB and KIBA, separately. The models are trained on the debiased datasets and evaluated on the test sets to measure the performance of AVE.

\section{Results}

\subsection{DebiasedDTA Improves Drug-Target Affinity Prediction Models}
We evaluate DebiasedDTA training framework with five different DTA prediction models (DeepDTA, BPE-DTA, LM-DTA, GraphDTA, and MGraphDTA) and with two different guides (BoW-DTA and ID-DTA). We train the models on BDB and KIBA and report test CI and R$^2$ in \Tabref{tab:bdb_debiasing} and \Tabref{tab:kiba_debiasing}, respectively, respectively. We also conduct a cross-dataset evaluation, whose details are available in the appendix.

\paragraph{The Overall Improvements due to DebiasedDTA} We first examine the performance increases due to DebiasedDTA and compare the best DebiasedDTA score on each setup with that score obtained without debiasing. The results show that at least one model trained in DebiasedDTA framework outperforms the non-debiased counterpart on 17 of 20  (85\%) evaluation setups of BDB (\Tabref{tab:bdb_debiasing}) and on 14 of 20 (70\%) evaluation setups of KIBA (\Tabref{tab:kiba_debiasing}). This highlights the success of the proposed training framework to enhance DTA prediction performance. The results also show that improvement in performance due to DebiasedDTA is more evident in the cold test sets of BDB, which present the most difficult generalization task due to the high diversity of BDB biomolecules. Although DebiasedDTA is mainly intended to improve DTA prediction performance on such difficult generalization setups, on all warm test setups at least one model trained in DebiasedDTA  improves the performance, too. This indicates that mitigating training set biases helps models to better predict affinities for known biomolecules as well.

Finally, \Tabref{tab:bdb_debiasing} and \Tabref{tab:kiba_debiasing} show that debiasing improves the performance of all affinity prediction models used in the study on at least one test setup. This emphasizes that DTA prediction models are susceptible to dataset biases irrespective of their input representation and the proposed training framework is general enough to mitigate biases in different biomolecule representation settings.

\paragraph{Benchmarking DebiasedDTA} In order to observe how DebiasedDTA compares to an existing method, we run every model on training sets debiased with AVE and compare the test set results.
The results show that in 13 of 20 test sets (65\%) of BDB (\Tabref{tab:bdb_debiasing}) and in 11 of 20 test sets (55\%) of KIBA (\Tabref{tab:kiba_debiasing}), at least one model trained via DebiasedDTA outperforms AVE debiasing in both metrics, while in 2 test sets of BDB and 6 test sets of KIBA AVE debiasing results in higher scores than DebiasedDTA training. These suggest that DebiasedDTA is a more successful approach than AVE, a state-of-the-art debiasing method proposed for ligand-based virtual screening, on the drug-target affinity prediction task.

\Tabref{tab:bdb_debiasing} and \Tabref{tab:kiba_debiasing} also demonstrate that AVE debiasing underperforms non-debiased model on 6 of 40 test setups in both metrics, aligning with the studies that claim existing debiasing approaches can lower prediction performance \citep{sundar2019effect}. DebiasedDTA, on the other hand, underperforms the non-debiased models with both guides only on 1  test setup and stands as a safer debiasing framework.

\renewcommand{\arraystretch}{1.35}
\begin{table*}
\caption{The effect of different debiasing strategies on the model performance per interaction types of the BDB dataset. We train each model 5 times using different folds of the training set and compute mean test set scores of the models. We report mean and standard deviation (in parentheses) of CI and R$^2$ metrics in the table. Mean squared errors and root mean squared errors, which are in parallel with R$^2$, are also available in the project repository.}
\begin{center}
\resizebox{\textwidth}{!}{
\begin{tabular}{|l|lcccccccc|}
\cline{1-10}
\multicolumn{1}{|c}{\textbf{\phantom{~~~}}}&&          \multicolumn{2}{c}{Warm}                &   \multicolumn{2}{c}{Cold Chemical}                &  \multicolumn{2}{c}{Cold Protein}                &     \multicolumn{2}{c|}{Cold Both}                \\ \cline{1-10}
\multicolumn{1}{|c}{\phantom{~}} & Debiasing & CI & R$^2$ & CI & R$^2$ & CI & R$^2$ & CI & R$^2$ \\ \cline{1-10}
\parbox[t]{.2mm}{\multirow{4}{*}{\rotatebox[origin=c]{90}{DeepDTA}}}& None & 0.888 (0.009) &  0.781 (0.028) & 0.687 (0.096) &  0.039 (0.243) & 0.759 (0.006) &  0.315 (0.049) & 0.554 (0.047) & -0.154 (0.164) \\
& AVE  & 0.883 (0.009) &  0.774 (0.025) & \textbf{0.704 (0.047)} &  -0.042 (0.023) & 0.765 (0.013) &  \textbf{0.368 (0.074)} & 0.565 (0.043) & -0.209 (0.150) \\ \cdashline{2-10}
& BoW-DTA  & \textbf{0.899 (0.004)} &  0.799 (0.013) & 0.698 (0.037) &  \textbf{0.043 (0.108)} & \textbf{0.777 (0.014)} &  0.351 (0.090) & 0.568 (0.044) & \textbf{-0.092 (0.132)} \\
& ID-DTA  & 0.898 (0.005) &  \textbf{0.804 (0.011)} & 0.693 (0.058) &  0.026 (0.109) & 0.771 (0.007) &  0.339 (0.067) & \textbf{0.585 (0.040)} & -0.128 (0.056) \\ \cline{1-10}

\parbox[t]{.2mm}{\multirow{4}{*}{\rotatebox[origin=c]{90}{BPE-DTA}}}& None & 0.883 (0.006) &  0.774 (0.013) & 0.657 (0.083) & -0.143 (0.202) & 0.653 (0.060) & \textbf{-0.256 (0.411)} & 0.522 (0.054) & -0.442 (0.349) \\
& AVE  & 0.881 (0.008) &  0.764 (0.016) & 0.667 (0.027) &  -0.075 (0.176) & 0.644 (0.032) &  -0.821 (0.267) & 0.536 (0.048) & -0.669 (0.627) \\  \cdashline{2-10}
& BoW-DTA  & 0.888 (0.008) &  \textbf{0.781 (0.016)} & 0.687 (0.082) & -0.091 (0.302) & \textbf{0.664 (0.067)} & -0.386 (0.593) & \textbf{0.568 (0.084)} & \textbf{-0.334 (0.347)} \\
& ID-DTA  & \textbf{0.891 (0.005)} &  0.777 (0.019) & \textbf{0.692 (0.065)} & \textbf{-0.045 (0.252)} & 0.650 (0.039) & -0.689 (0.476) & 0.565 (0.090) & -0.426 (0.231) \\ \cline{1-10}
 
\parbox[t]{.2mm}{\multirow{4}{*}{\rotatebox[origin=c]{90}{LM-DTA}}}& None & 0.876 (0.005) & 0.745 (0.011) & 0.688 (0.046) & -0.027 (0.175) & 0.780 (0.016) &  0.384 (0.083) & 0.572 (0.028) & -0.226 (0.205) \\
& AVE  & 0.880 (0.018) &  0.743 (0.019) & \textbf{0.705 (0.038)} &  -0.008 (0.234) & \textbf{0.785 (0.018)} &  \textbf{0.396 (0.080)} & \textbf{0.592 (0.033)} & -0.183 (0.242) \\  \cdashline{2-10}

& BoW-DTA  & 0.882 (0.006) &\textbf{0.762 (0.003)} & 0.688 (0.069) & \textbf{-0.005 (0.169)} & 0.781 (0.017) &  0.386 (0.081) & 0.563 (0.032) & \textbf{-0.182 (0.136)} \\

& ID-DTA  & \textbf{0.883 (0.006)} & 0.758 (0.003) & 0.683 (0.067) & -0.016 (0.270) & 0.782 (0.017) &  0.387 (0.080) & 0.581 (0.017) & -0.198 (0.174) \\ \cline{1-10}

\parbox[t]{.2mm}{\multirow{4}{*}{\rotatebox[origin=c]{90}{GraphDTA}}}& None & 0.824 (0.010) & 0.493 (0.060) & 0.701 (0.024) &	\textbf{0.143 (0.138)} &	0.685 (0.039) &	0.040 (0.114)	& 0.558 (0.077)	& -0.047 (0.162) \\

& AVE  & 0.825 (0.017) & 0.502 (0.072) & 0.716 (0.049) & 0.094 (0.186) & 0.692 (0.027) &	0.029 (0.073) &	0.581 (0.082) &	-0.057 (0.143)
 \\  \cdashline{2-10}

& BoW-DTA  & \textbf{0.835 (0.015)} & \textbf{0.511 (0.078)} & 0.706 (0.032) & 0.085 (0.091) & 0.690 (0.034) & 0.045 (0.062) & 0.556 (0.065) & -0.061 (0.118) \\

& ID-DTA  & 0.832 (0.012) & 0.507 (0.077) & \textbf{0.728 (0.039)} & 0.090 (0.102) & \textbf{0.695 (0.027)} & \textbf{0.049 (0.072)} & \textbf{0.603 (0.050)} & \textbf{-0.018 (0.071)}
 \\ \cline{1-10}

\parbox[t]{.2mm}{\multirow{4}{*}{\rotatebox[origin=c]{90}{MGraphDTA}}}& None&0.834 (0.013)&0.520 (0.067)&0.684 (0.030)&-0.125 (0.134)&0.754 (0.013)&0.289 (0.070)&0.555 (0.059)&-0.448 (0.497) \\
&AVE&\textbf{0.837 (0.013)}&0.509 (0.073)&0.687 (0.042)&-0.109 (0.125)&\textbf{0.758 (0.017)}&\textbf{0.295 (0.075)}&0.551 (0.031)&-0.429 (0.386) \\ \cdashline{2-10}
&BoW-DTA&\textbf{0.837 (0.013)}&\textbf{0.522 (0.078)}&\textbf{0.705 (0.029)}&\textbf{-0.040 (0.168)}&0.748 (0.024)&0.279 (0.089)&\textbf{0.580 (0.042)}&\textbf{-0.313 (0.353)} \\
&ID-DTA&\textbf{0.837 (0.014)}&0.521 (0.076)&0.698 (0.036)&-0.266 (0.175)&0.755 (0.018)&0.281 (0.074)&0.574 (0.054)&-0.606 (0.588)  \\ \cline{1-10}

\end{tabular}
}
\end{center}
\label{tab:bdb_debiasing}
\end{table*}

\renewcommand{\arraystretch}{1}
\renewcommand{\arraystretch}{1.35}
\begin{table*}
\caption{The effect of different debiasing strategies on the model performance per interaction type of the KIBA dataset. We train each model 5 times using different folds of the training set and compute mean test set scores of the models. We report mean and standard deviation (in parentheses) of CI and R$^2$ metrics in the table. Mean squared errors and root mean squared errors, which are in parallel with R$^2$, are also available in the project repository.}
\begin{center}
\resizebox{\textwidth}{!}{
\begin{tabular}{|l|lcccccccc|}
\cline{1-10}
\multicolumn{1}{|c}{\phantom{~~~}}&&          \multicolumn{2}{c}{Warm}                &   \multicolumn{2}{c}{Cold Chemical}                &  \multicolumn{2}{c}{Cold Protein}                &     \multicolumn{2}{c|}{Cold Both}                \\ \cline{1-10}
\multicolumn{1}{|c}{\phantom{~}} & Debiasing & CI & R$^2$ & CI & R$^2$ & CI & R$^2$ & CI & R$^2$ \\ \cline{1-10}

\parbox[t]{.2mm}{\multirow{4}{*}{\rotatebox[origin=c]{90}{DeepDTA}}}& None & 0.873 (0.005) &  0.756 (0.021) & 0.753 (0.018) &  0.337 (0.081) & 0.719 (0.029) &  0.330 (0.109) & 0.654 (0.019) &  \textbf{0.087 (0.099)} \\
& AVE  & 0.874 (0.004) &  0.747 (0.018) & 0.754 (0.014) &  \textbf{0.351 (0.050)} & 0.713 (0.044) &  \textbf{0.343 (0.113)} & 0.643 (0.033) & 0.081 (0.176) \\  \cdashline{2-10}
& BoW-DTA  &\textbf{0.888 (0.005)} &  \textbf{0.775 (0.019)} & \textbf{0.761 (0.004)} &  0.349 (0.046) & 0.713 (0.036) &  0.308 (0.115) & 0.639 (0.028) &  0.045 (0.147) \\
& ID-DTA  & 0.887 (0.006) &  \textbf{0.775 (0.018)} & \textbf{0.761 (0.020)} &  0.350 (0.101) & \textbf{0.725 (0.038)} &  0.333 (0.124) & \textbf{0.660 (0.034)} &  0.084 (0.195) \\ \cline{1-10}
 
\parbox[t]{.2mm}{\multirow{4}{*}{\rotatebox[origin=c]{90}{BPE-DTA}}}& None & 0.881 (0.005) &  0.760 (0.016) & 0.735 (0.025) &  \textbf{0.274 (0.105)} & 0.680 (0.020) &  0.185 (0.077) & \textbf{0.605 (0.033)} & \textbf{-0.006 (0.117)} \\
& AVE  &  0.883 (0.004) &  0.754 (0.022) &  \textbf{0.736 (0.029)} &  0.256 (0.139) & \textbf{0.689 (0.032)} &  \textbf{0.222 (0.114)} & 0.576 (0.020) & -0.117 (0.070) \\ \cdashline{2-10}
& BoW-DTA  & 0.891 (0.003) &  0.774 (0.016) & \textbf{0.736 (0.018)} &  0.231 (0.093) & 0.679 (0.030) &  0.174 (0.103) & 0.604 (0.017) & -0.046 (0.082) \\
& ID-DTA  & \textbf{0.893 (0.003)} &  \textbf{0.776 (0.012)} & \textbf{0.736 (0.021)} &  0.229 (0.099) & 0.684 (0.023) &  0.179 (0.060) & 0.590 (0.014) & -0.037 (0.079) \\ \cline{1-10}
 
\parbox[t]{.2mm}{\multirow{4}{*}{\rotatebox[origin=c]{90}{LM-DTA}}}& None & 0.858 (0.005) & 0.756 (0.012) & 0.749 (0.012) &  0.409 (0.067) & 0.713 (0.049) &  0.366 (0.137) & 0.650 (0.041) &  0.107 (0.122) \\
& AVE  &  0.861 (0.005) &  0.758 (0.016) &   0.754 (0.012) &  0.417 (0.065) & 0.713 (0.052) &  0.381 (0.014) & 0.649 (0.026) & 0.135 (0.108) \\  \cdashline{2-10}
& BoW-DTA  & \textbf{0.865 (0.005)} & \textbf{0.769 (0.013)} & 0.756 (0.013) &  0.435 (0.064) & 0.717 (0.051) &  0.382 (0.139) & \textbf{0.653 (0.028)} &  \textbf{0.159 (0.121)} \\
& ID-DTA  & 0.864 (0.006) & 0.767 (0.014) & \textbf{0.759 (0.011)} &  \textbf{0.436 (0.056)} & \textbf{0.718 (0.053)} &  \textbf{0.385 (0.143)} & 0.652 (0.036) &  0.151 (0.126) \\ \cline{1-10}

\parbox[t]{.2mm}{\multirow{4}{*}{\rotatebox[origin=c]{90}{GraphDTA}}}& None & 0.882 (0.006) & 0.782 (0.008) & 0.765 (0.022) & 0.418 (0.078) & 0.663 (0.039) & \textbf{0.247 (0.093)} & 0.604 (0.038) & \textbf{0.081 (0.086)} \\

& AVE  & 0.881 (0.006) & 0.772 (0.024) & 0.771 (0.017) & \textbf{0.431 (0.052)} & \textbf{0.684 (0.053)}& 0.245 (0.078) & \textbf{0.623 (0.054)} & 0.071 (0.089) \\  \cdashline{2-10}

& BoW-DTA  & 0.885 (0.005) & 0.792 (0.015) & 0.767 (0.015) & 0.425 (0.072) & 0.666 (0.043) & 0.220 (0.062) & 0.619 (0.024) & 0.047 (0.060) \\

& ID-DTA  & \textbf{0.887 (0.006)} & \textbf{0.797 (0.017)} & \textbf{0.776 (0.009)} &0.429 (0.043) & 0.680 (0.045) &	0.241 (0.056) & 0.617 (0.020) &	0.048 (0.058) \\ \cline{1-10}

\parbox[t]{.2mm}{\multirow{4}{*}{\rotatebox[origin=c]{90}{MGraphDTA}}}&None&0.899 (0.004)&0.807 (0.015)&0.750 (0.021)&0.306 (0.103)&0.712 (0.044)&0.361 (0.115)&0.618 (0.033)&-0.023 (0.218) \\ 
&AVE&0.900 (0.004)&0.809 (0.017)&\textbf{0.762 (0.018)}&\textbf{0.378 (0.088)}&0.726 (0.055)&0.385 (0.118)&\textbf{0.624 (0.052)}&\textbf{0.059 (0.173)} \\ \cdashline{2-10}
&BoW-DTA&0.900 (0.006)&0.808 (0.019)&0.752 (0.016)&0.343 (0.090)&0.717 (0.054)&0.376 (0.112)&0.622 (0.049)&0.014 (0.181) \\ 
&ID-DTA&\textbf{0.901 (0.006)}&\textbf{0.811 (0.020)}&0.759 (0.025)&0.344 (0.129)&\textbf{0.736 (0.045)}&\textbf{0.398 (0.110)}&0.622 (0.040)&0.007 (0.226) \\ \cline{1-10}
\end{tabular}
}
\end{center}
\label{tab:kiba_debiasing}
\end{table*}
\renewcommand{\arraystretch}{1}

\subsection{DebiasedDTA Enhances Learning from Proteins}

The experiments show that DebiasedDTA can improve DTA prediction models with different biomolecule representations on similar and distant test sets. Here, we investigate how DebiasedDTA affects the relationship between inputs and model predictions. The setup of ID-DTA and DeepDTA is arbitrarily selected and re-trained on an arbitrary setup of BDB dataset. The contribution of each DeepDTA  feature (biomolecule characters) to the warm test set predictions is examined via Gradient-weighted Class Activation Mapping  GradCAM \citep{gradcam}. Given a protein-chemical pair and the model prediction, GradCAM outputs an "attention coefficient" that quantifies the contribution of each input feature to the prediction. We run GradCAM on debiased and non-debiased DeepDTA models and acquire attention coefficients of each feature for warm test set predictions.

We compare the average attention coefficients of protein and chemical characters for each test set interaction since inadequate learning from proteins is a known challenge for having generalizable DTA models \citep{wallach2018most,chen2019hidden,sieg2019need,scantlebury2020data}. The comparison shows for non-debiased DeepDTA that in 65\% of the predictions, the most attended feature is a chemical character, while for the debiased DeepDTA the same statistic is computed as 56\%. This finding indicates that DebiasedDTA pushes models to attribute more importance to protein characters and is a step towards learning more from the proteins to achieve generalizability.

\section{Conclusion}
DebiasedDTA is a model training framework that aims to improve the generalization performance of DTA prediction models, overcoming problems created by dataset biases. The problem of out-of-distribution generalization is a notoriously difficult one, especially so for DTA prediction tasks. Our methodology and results point to a promising direction for enhancing the toolbox for  generalization in DTA prediction. %
The framework we propose can be used by other researchers to develop, integrate, and benchmark their own guide and/or predictor methods to contribute to generalization research in DTA prediction.

Future directions for our research include improvements on the proposed method and further investigation of the nature of the problem. An understanding of how our method and its hyperparameters interact with various other models \textit{and} datasets is crucial. A more detailed statistical analysis of our weighting scheme to further understand how it affects the estimation task in relation to the specifics of the training distribution would also be very illuminating. Finally, applying our method to non-DTA prediction tasks would help further understand what makes DTA datasets and tasks unique; these observations can then be leveraged to further improve DTA predictions.

\section*{Acknowledgements}
This work was supported by the Scientific and Technological Research Council of Turkey [Grant Number 119E133]. We thank Selen Parlar for her contributions to Figure 1.

\bibliographystyle{abbrvnat}
\bibliography{references}
\appendix
\newpage
\section{Supplementary Material}
\begin{table}
\centering
\caption{Dataset statistics. Mean number of proteins, chemicals, and interactions per test set type for both datasets are reported. Standard deviations are also provided next to the means. C. stands for ``cold" in the table.}
\begin{tabular}{|l|llll|}
\hline
\multicolumn{1}{|c}{\phantom{~~~}} & 
Fold & \# Proteins & \# Chemicals & \# Interactions \\ \hline

\parbox[t]{.2mm}{\multirow{6}{*}{\rotatebox[origin=c]{90}{BDB}}} & Train & 403.4 $\pm$ 2.8 & 740.8 $\pm$ 19.46 & 17988.2 $\pm$ 646.45 \\
& Validation & 355.0 $\pm$ 5.62 & 170.0 $\pm$ 11.05 & 1494.2 $\pm$ 56.17 \\
& Warm       & 354.4 $\pm$ 3.44 & 179.6 $\pm$ 5.28    & 1494.4 $\pm$ 56.32   \\
& C. Chemical  & 376.0 $\pm$ 4.38 & 84.8 $\pm$ 5.53     & 2448.8 $\pm$ 373.48  \\
& 
C. Protein & 43.6 $\pm$ 2.15  & 264.8 $\pm$ 90.17   & 2360.0 $\pm$ 216.02  \\
& 
C. Both & 41.4 $\pm$ 3.07  & 30.8 $\pm$ 11.92    & 274.6 $\pm$ 36.19    \\ \hline
\parbox[t]{.2mm}{\multirow{6}{*}{\rotatebox[origin=c]{90}{KIBA}}} & Train & 200.6 $\pm$ 1.36 & 1834.6 $\pm$ 6.41 & 77264.4 $\pm$ 814.94 \\
& Validation & 193.0 $\pm$ 1.67 & 1467.2 $\pm$ 23.75 & 6650.2 $\pm$ 69.53 \\
& Warm       & 192.0 $\pm$ 3.16 & 1476.2 $\pm$ 17.7   & 6650.6 $\pm$ 69.1    \\
& 
C. Chemical  & 193.0 $\pm$ 2.45 & 140.0 $\pm$ 5.59    & 6810.0 $\pm$ 570.52  \\
& 
C. Protein & 14.6 $\pm$ 0.8   & 1296.0 $\pm$ 179.09 & 6259.6 $\pm$ 1024.25 \\
& 
C. Both & 14.0 $\pm$ 1.1   & 100.2 $\pm$ 14.55   & 468.6 $\pm$ 37.89  \\
\hline
\end{tabular}
\label{tab:dataset_stats}
\end{table}

\subsection{Dataset Setup Creation Details}

 We evaluate the models with five distinct train-test setups per dataset. To create different setups, we cluster the proteins and chemicals in the datasets and randomly divide the clusters  into two as ``warm" and ``cold" via hierarchical clustering with ward linkage. To create the pairwise distance matrices required for hierarchical clustering, we computed the Tanimoto similarity  between Morgan fingerprints of compounds and subtracted the result from 1. For proteins, we used the rows of normalized Smith-Waterman score matrix for representation and subtract the elements of the matrix from 1 to compute the distance matrix. The ward linkage clustering suggested 800 hundred clusters for compounds and 400 clusters for proteins.
 We interpret the warm clusters as previously encountered biomolecules and the cold clusters as novel or unseen biomolecules. To produce training and test sets from warm and cold biomolecule clusters, we first select the interactions between proteins and chemicals in the warm clusters. We use these interactions mainly as the training set, but also separate small subsets as ``validation" and ``warm test" sets. The validation fold is used to tune model hyper-parameters, whereas the warm test set is utilized to evaluate models on the interactions between seen biomolecules.
We create two more test sets called ``cold chemical" and ``cold protein", where the cold chemical test set consists of the interactions between chemicals in the cold cluster and proteins in the warm cluster. This test set is used to estimate model performance when new drugs are searched to target existing proteins. The cold protein test set is created similarly and used to evaluate models in the scenarios where existing drugs are searched to target a novel protein.   Last, we create a ``cold both" test set, which is the set of interactions between the proteins and chemicals in the cold clusters. This is the most challenging test set of every setup, as both the proteins and the chemicals are unavailable in the training set. The average number of proteins, chemicals, and interactions in the training and test sets are reported in \Tabref{tab:dataset_stats}, alongside standard deviations.

\subsection{Cross Dataset Evaluation}
Experiments on BDB and KIBA show that DebiasedDTA can enhance the generalizability of the DTA prediction models, especially when the test set has a significantly different distribution, as in the cold both setups of the BDB dataset. Here, we further challenge the proposed methodology by out-of-dataset interaction predictions: we use the models trained on BDB to predict the affinity of all protein-chemical pairs in KIBA, and vice versa. Prior to prediction, we remove the SMILES-amino acid sequence pairs shared between the datasets to eliminate the risk of information leak from the test set to the training set, and binarize the labels (strong binding/weak binding) since BDB and KIBA report the affinity scores in terms of inconvertible metrics. \Tabref{tab:cross} reports the mean and standard deviation of the F1 score for the non-debiased model and, for brevity, the best debiased model.

\Tabref{tab:cross} demonstrates that DebiasedDTA achieves a higher mean cross-dataset F1-score than the non-debiased models in 8 of 10 test setups, indicating that DebiasedDTA can improve prediction on out-of-dataset interactions, too, arguably the most challenging generalization task.

Another finding in \Tabref{tab:cross} is that LM-DTA trained on BDB dataset achieves the highest  scores across setups by a landslide. We relate this to the pre-training phases of ChemBERTa and ProtBERT models used by LM-DTA. During the pre-training, these models are exposed to thousands of kinase family proteins and their ligands and this may improve the generalizability of LM-DTA to a kinase dataset, KIBA.

\begin{table}
\caption{Binary evaluation of the models on cross-dataset. We use the previously learned weights for each model and predict affinity of the cross-dataset interactions. We convert the predicted and reported affinity scores to binary labels and measure F1-scores. Mean and standard deviation (in parentheses) of 5 different weights for each model are reported.}
\label{tab:cross}
\centering
\begin{tabular}{|l|lcc|}
\cline{1-4}
\multirow{1}{*}{Training Dataset} & Model\hspace{7.5em} & \hspace{0em}No Debiasing\hspace{0em} & \hspace{0em}DebiasedDTA\hspace{0em} \\  \cline{1-4}

\multirow{5}{*}{BDB} & DeepDTA     & 0.146 (0.025)  &  0.152 (0.011) \\

& BPE-DTA     & 0.168 (0.040) & 0.186 (0.042)  \\

& LM-DTA      & 0.520 (0.031) & 0.522 (0.021) \\

& GraphDTA      & 0.152 (0.022)  & 0.155 (0.015) \\

& MGraphDTA      & 0.241 (0.011)  & 0.233 (0.019) \\  \cline{1-4}

\multirow{5}{*}{KIBA} & DeepDTA     &  0.246 (0.021) & 0.243 (0.037) \\ 

& BPE-DTA     &  0.190 (0.040) & 0.217 (0.018) \\

& LM-DTA      &  0.286 (0.019) & 0.289 (0.016) \\ 

& GraphDTA   & 0.324 (0.006)  & 0.327 (0.015) \\

& MGraphDTA      & 0.330 (0.010) & 0.339 (0.006) \\ \cline{1-4}
\end{tabular}
\end{table}

\end{document}